\documentclass{amsart}
\newtheorem{theorem}{Theorem}[section]
\newtheorem{lemma}[theorem]{Lemma}

\theoremstyle{definition}

\theoremstyle{remark}

\numberwithin{equation}{section}

\def\t{\tau}
\def\il{\int\limits}
\def\g{\gamma}
\def\f{\frac}
\def\q{\quad}
\def\p{\varphi}
\def\r{\rho}
\def\P{\Phi}
\def\l{\lambda}
\def\L{\Lambda}

\def\b{\beta}
\def\s{\sigma}
\def\ep{\varepsilon}

\def\ba{\begin{array}}
\def\pr{\prime}

\begin{document}

\begin{center}
{\Large\bf Continuous regularized Gauss-Newton-type algorithm for nonlinear}\\
{\Large\bf ill-posed equations with simultaneous updates of inverse derivative}
\\

\vspace{0.5cm}

Alexander G. Ramm\\
E-mail: ramm@math.ksu.edu\\
Department of Mathematics\\
Kansas State University\\Manhattan, KS 66506, U.S.A.\\
\vspace{0.3cm}

Alexandra B. Smirnova\\
E-mail: smirn@cs.gsu.edu\\
Department of Mathematics and Statistics\\
Georgia State University\\Atlanta, GA 30303, U.S.A.\\
\end{center}
\vspace{2cm}

\noindent {\it A new continuous regularized Gauss-Newton-type
method with simultaneous updates of the operator
$(F^{\pr*}(x(t))F'(x(t))+\ep(t) I)^{-1}$ for solving nonlinear
ill-posed equations in a Hilbert space is proposed. A
convergence theorem is proved. An attractive and novel feature of the
proposed  method  is the absence of the assumptions about the
location of the spectrum of the operator $F'(x)$. The absence of
such assumptions is made possible by a source-type condition.}
\vspace{5mm}


\vspace{0.3cm}

\section{{\bf Introduction}}
\setcounter{equation}{0} \setcounter{theorem}{0}
\renewcommand{\thetheorem}{1.\arabic{theorem}}
\renewcommand{\theequation}{1.\arabic{equation}}
\vspace*{-0.5pt}

Consider a nonlinear operator equation
\begin{equation}\label{1.1}
F(x)=0,\q F: H\to H,
\end{equation}
in a real Hilbert space $H$. We suppose here that equation
(\ref{1.1}) is solvable (not necessarily uniquely), and the
operator $F$ is twice Fr\'echet differentiable  without such
structural assumptions as monotonicity, invertibility of $F'$ etc.

In the theory of ill-posed problems various discrete and
continuous methods based on a regularization are known. A
principal point in the numerical implementation of regularized
Newton's and Gauss-Newton's procedures is the inversion of the
operators $(F'(x)+\ep I)$ and $(F^{\pr*}(x)F'(x)+\ep I)$
respectively \cite{ars1}, \cite{ars2}, \cite{bns}. Such an
inversion for certain operators is a nontrivial task. Moreover,
the inversion decreases the accuracy of computations.

For well-posed equations (i.e., in the case when the Fr\'echet
derivative operator $F'(x)$ is boundedly invertible in a ball,
which contains one of the solutions) several approaches are taken
in order to reduce the cost associated with the storage and
inversion of $F'(x)$ in Newton's scheme (or $F^{\pr*}(x)F'(x)$ in
Gauss-Newton's scheme). Iterative techniques for linear systems,
like 'conjugate gradient', can be applied to compute an
approximation to the Newton step $s_n:=x_{n+1}-x_n$, yielding an
inexact, or truncated, Newton's method. Such a method does not
explicitly require the operator $F'(x)$. Instead, it requires only
applications of $F'(x)$ to certain elements. See \cite{des},
\cite{s} or (\cite{nw}, p.136).

Another approach is to replace the exact Fr\'echet derivative by an
approximation obtained from current and previous values of $F(x)$.
In this case, only $F(x_{n+1})$ and $F(x_n)$ are required. This
generates the secant method (\cite{ds}, p.201). Perhaps the most
popular secant method is the BFGS method, which was discovered
independently by Broyden, Fletcher, Goldfarb, and Shanno in 1970.
To implement the BFGS, given an approximation $J_n$ to $F'(x_n)$,
one first computes
$$
s_n:=x_{n+1}-x_n \q \mbox{and} \q y_n:=F(x_{n+1})-F(x_n).
$$
One then gets an approximation to $F'(x_{n+1})$:
$$
J_{n+1}=J_n-\f{(J_ns_n)(J_ns_n)^T}{s_n^TJ_ns_n}+\f{y_ny_n^T}{y_n^Ts_n}.
$$
If $J_n$ is symmetric positive definite and $y_n^Ts_n>0$, then
$J_{n+1}$ will also be symmetric positive definite. In the finite
dimensional well-posed case, under standard assumptions, the BFGS
method is guaranteed to be superlinearly convergent.

BFGS has a limited memory variant (\cite{nw}, p.224), which
requires no explicit matrix storage for the approximate
derivative. It is based on the recursion for the inverse
$$
J^{-1}_{n+1}=\left(I-\f{s_ny_n^T}{y_n^Ts_n}\right)J^{-1}_n\left(I-\f{y_ns_n^T}{y_n^Ts_n}\right)+\f{s_ns_n^T}{y_n^Ts_n}.
$$
In \cite{a} the following continuous Newton-type algorithm for
solving nonlinear well-posed operator equations is investigated
$$
\dot x(t)=-J(t)F(x(t)),\q x(0)=x_0 \in H, \q J(0)\in L(H),\hskip
80 pt
$$
$$
\dot J(t)=-\r^2\bigl[F^{\pr*}(x(t))F'(x(t))J(t)+
J(t)F'(x(t))F^{\pr*}(x(t))\bigr]+2\r^2F^{\pr*}(x(t)).
$$
The theorem establishing convergence with the exponential rate is
proved for the above procedure.

In many important applications the Fr\'echet derivative operator
is not boundedly invertible: this is an ill-posed case. In order
to deal with this difficulty, we propose a novel continuous
algorithm with simultaneous updates of
$(F^{\pr*}(x(t))F'(x(t))+\ep(t) I)^{-1}$ and prove a convergence
theorem. The paper is organized as follows. In section 2 we
introduce the algorithm and also state two Lemmas. In section 3
the main convergence result is established. An attractive feature
of this result (formulas (\ref{2.3})-(\ref{2.4}) in section 2) is
the absence of the assumptions about the location of the spectrum
of the operator $F'(x)$. For example, in \cite{ar} the operator
$F'(x)$ was assumed nonnegative, which is a strong assumption
about location of its spectrum. One may generalize the methods and
results of \cite{ar} to the case of the operators for which $\Re
F'(x) \geq 0$, which still imposes a restriction on the location
of the spectrum of $F'(x)$. The absence of such assumptions in our
paper is made possible by the conditions 5) and 6) of
Theorem~\ref{t1} in section 3.

\section{{\bf The Initial Value Problem and Some Auxiliary Results}}
\setcounter{equation}{0} \setcounter{theorem}{0}
\renewcommand{\thetheorem}{2.\arabic{theorem}}
\renewcommand{\theequation}{2.\arabic{equation}}
\vspace*{-0.5pt}

Consider continuously regularized Gauss-Newton's procedure for
solving equation (\ref{1.1}):
\begin{equation}\label{2.1}
\dot
x(t)=-\bigl[F^{\pr*}(x(t))F'(x(t))+\ep(t)I\bigr]^{-1}\bigl[F^{\pr*}(x(t))F(x(t))+\ep(t)(x(t)-x_0)\bigr],
\end{equation}
where $\ep(t)>0$, $x_0\in H$, and $I$ is an identity operator. The
reader may consult \cite{ars1} and \cite{ars2} for a convergence
analysis of (\ref{2.1}). Problem (\ref{2.1}) is equivalent to the
following system:
$$
\dot
x(t)=-Q(t)\bigl[F^{\pr*}(x(t))F(x(t))+\ep(t)(x(t)-x_0)\bigr],\hskip
28pt
$$
\begin{equation}\label{2.2}
[F^{\pr*}(x(t))F'(x(t))+\ep(t) I]Q(t)-I=0,\q Q(t)\in L(H).
\end{equation}
Solve equation (\ref{2.2}) by using continuous simple iteration
scheme. We arrive at the initial value problem
\begin{equation}\label{2.3}
\dot x(t)=-B(t)\bigl[F^{\pr*}(x(t))F(x(t))+\ep(t)(x(t)-x_0)\bigr],
\end{equation}
\begin{equation}\label{2.4}
\dot B(t)=-\bigl[(F^{\pr*}(x(t))F'(x(t))+\ep(t)
I)B(t)-I\bigr],\hskip 13pt
\end{equation}
$$
x(0)=x_0\in H,\q B(0)\in L(H),\q 0<\ep(t)\to 0\q \mbox{as}\q t\to
+\infty.
$$

The following lemma about the inversion of a nonlinear
differential inequality was first stated and proved in
\cite{ars2}.

\begin{lemma}
Let $\g(t), \s(t), \b(t)\in C[0,+\infty)$. If there exists a
positive  function $\mu(t)\in C^1[0,+\infty)$ such that
\begin{equation}\label{ric1}
0\le\s(t)\le\f{\mu(t)}{2}\left(\g(t)-\f{\dot{\mu}(t)}{\mu(t)}\right),\q
\b(t)\le\f{1}{2\mu(t)}\left(\g(t)-\f{\dot{\mu}(t)}{\mu(t)}\right),\q
\mu(0)v(0)<1,
\end{equation}
then a nonnegative solution to the following inequality
\begin{equation}\label{ric4}
\dot{v}(t)\le -\g(t) v(t)+\s(t)v^2(t)+\b(t)
\end{equation}
satisfies the estimate:
\begin{equation}\label{ric5}
0\,\leq v(t)\,<\, \f{1}{\mu(t)}.
\end{equation}
\label{l2}\end{lemma}

Now, we formulate the second lemma, which is an
operator-theoretical version of the well-known Gronwall
inequality:

\begin{lemma}
Let
\begin{equation}\label{2.5}
\frac{d V}{d t} + A(t)V(t)=G(t),\q V(0)=V_0,
\end{equation}
where $A(t),$ $G(t),$ $V(t)\in L(H),$ and $H$ is a real Hilbert
space. If there exists $\g(t)>0$ such that
\begin{equation}\label{2.6}
(A(t)h,h)\ge \g(t)||h||^2 \q \forall h\in H,
\end{equation}
then
\begin{equation}\label{2.7}
||V(t)||\le
e^{-\il^t_0\g(p)dp}\left[\il^t_0||G(s)||e^{\il^s_0\g(p)dp}\,ds+||V(0)||\right].
\end{equation}
\label{l3}\end {lemma}

{\bf Proof.} Take any $h\in H.$ Since $H$ is a real Hilbert space
one has:
$$
\hskip -1.7cm \frac{1}{2} \frac{d}{dt}
||V(t)h||^2=\left(\frac{dV}{dt}h,V(t)h\right)
$$
$$
\hskip 4cm =-(A(t)V(t)h, V(t)h)+(G(t)h,V(t)h)
$$
\begin{equation}\label{2.8}
\hskip 117pt \le -\g(t)||V(t)h||^2+||G(t)||\,\, ||h||\,\,||V(t)h||
\end{equation}
Denote $v(t):=||V(t)h||$. Inequality (\ref{2.8}) implies
\begin{equation}\label{2.9}
v\dot v\le-\g(t)v^2+||G(t)||\,\,||h||\,v.
\end{equation}
Divide this inequality by the nonnegative $v$ and get a linear
first-order differential inequality from which one can derive the
desired inequality (\ref{2.7}). Lemma 2.3 is proved. $\qed$

\section{{\bf The Convergence Theorem}}
\setcounter{equation}{0} \setcounter{theorem}{0}
\renewcommand{\thetheorem}{3.\arabic{theorem}}
\renewcommand{\theequation}{3.\arabic{equation}}
\vspace*{-0.5pt}

\begin{theorem}
Let $H$ be a real Hilbert space, $F: H\to H$.

\noindent 1)Assume that $0<\ep(t)\in C^1[0,+\infty)$ converges to
zero monotonically and
\begin{equation}\label{ep}
|\dot\ep(t)|\le b \ep^2(t) \q \mbox{for any} \q t\in [0,+\infty),
\end{equation}
where $b>0$ is a constant.

\noindent 2) Suppose problem (\ref{1.1}) is solvable (not
necessarily uniquely) and $\hat x$ is its solution.

\noindent 3) There exists a positive number $R$ such that $F$ is
twice Fr\'echet differentiable in a closed ball $U(\hat
x,R\ep(0)):=\bigl\{x:\, x\in H,\,||x-\hat x||\le R\ep(0)\bigr\}$
and
$$
||F'(x)||\le N_1,\q ||F''(x)||\le N_2\q \mbox{for all}\q x\in
U(\hat x,R\ep(0)).
$$

\noindent 4) Let the following inequality be satisfied
\begin{equation}\label{?}
k+b\ep(0)<1,
\end{equation}
where $ k:=2N_1N_2R+b+\ep(0)||B(0)||+||I-B(0)\bigl[F^{\pr*}(\hat
x)F'(\hat x)+\ep(0)I\bigr]||. $

\noindent 5) For an initial approximation point $x_0$ the source-type
condition holds
\begin{equation}\label{s}
x_0\in U(\hat x,R\ep(0))\cap \bigl[\hat x+Ran(F^{\pr*}(\hat
x)F'(\hat x))\bigr],
\end{equation}
where $Ran(F^{\pr*}(\hat x)F'(\hat x))$ is the range of the linear
operator $F^{\pr*}(\hat x)F'(\hat x)$, and
 for some $w$, such that $\hat x-x_0=F^{\pr*}(\hat
x)F'(\hat x)w$, one has:
\begin{equation}\label{??}
\f{1}{R}\le\f{3N_1N_2(1+\ep(0)||B(0)||)}{1-k-b\ep(0)}<\min
\left\{\f{1-k-b\ep(0)}{2(k+2+\ep(0)||B(0)||)||w||},\,\,\f{\ep(0)}{||x(0)-\hat
x||}\right\}.
\end{equation}
Then

\noindent the solution $(x(t),B(t))$ to problem
(\ref{2.3})-(\ref{2.4}) exists for all $t\in [0,+\infty)$ and
\begin{equation}\label{lim}
||x(t)-\hat x||<R\ep(t).
\end{equation}
\label{t1}\end{theorem}

\noindent {\bf Remark 3.2} There are many $\ep(t)$ satisfying
condition 1) of Theorem 3.1. For example, one may take
$\ep(t)=c_0(c_1+t)^{-a}$, where $c_0$ and $c_1$ are positive
constants and $a\in (0,1] $.

\noindent {\bf Remark 3.3} Note that assumption (\ref{s}) is not
algorithmically verifiable. However, practitioners may try
different $x_0$ and choose the one for which the algorithm works
better, that is, convergence is more rapid and the algorithm is
more stable.

If $F^{\pr*}(\hat x)F'(\hat x)$ is compact and the null space
$N(F^{\pr*}(\hat x)F'(\hat x))=\{0\}$, then the range
$Ran(F^{\pr*}(\hat x)F'(\hat x))$ is dense in $H$, so in any
neighborhood of $\hat x$ there are points $x_0$ for which
(\ref{s}) holds. On the other hand, since
$F^{\pr*}(\hat x)F'(\hat x)$ is compact, the set
 $R(F^{\pr*}(\hat
x)F'(\hat x))$ is not closed. Thus, in the same neighborhood there are
also points $x_0$ for which (\ref{s}) fails to hold. In general,
in order to get a convergence theorem in an ill-posed case one
needs some additional assumptions on the Fr\'echet derivative of
the operator $F$, for example condition (\ref{s}), or some other
condition of this type ( see e.g., \cite{des1}, condition (2.11)).

\noindent {\bf Remark 3.4} By the assumption 4) of Theorem~\ref{t1}
the first inequality in (\ref{??}) is equivalent to the following
one:
\begin{equation}\label{r0}
3N_1N_2R(1+\ep(0)||B(0)||)\ge
1-2N_1N_2R-b-\ep(0)||B(0)||-||\L(0)||-b\ep(0),
\end{equation}
where $||\L(0)||:=|| I-B(0)\bigl[F^{\pr*}(\hat x)F'(\hat
x)+\ep(0)I\bigr]||$. Inequality (\ref{r0}) yields
\begin{equation}\label{r1}
R\ge
\f{1-b-\ep(0)||B(0)||-||\L(0)||-b\ep(0)}{(5+3\ep(0)||B(0)||)N_1N_2}.
\end{equation}
If one takes
\begin{equation}\label{r2}
R:=
\f{1-b-\ep(0)||B(0)||-||\L(0)||-b\ep(0)}{(5+3\ep(0)||B(0)||)N_1N_2},
\end{equation}
then condition (\ref{?}) is equivalent to
\begin{equation}\label{r3}
b+||\L(0)||+b\ep(0)[1+||B(0)||]+\ep(0)||B(0)||[\ep(0)||B(0)||
+||\L(0)||+\ep(0)]<1.
\end{equation}
For inequality (\ref{r3}) to hold one has to have $b$,
$||\L(0)||$ and $\ep(0)||B(0)||$ sufficiently small. The second
inequality in (\ref{??}) holds if $||w||$ and
$||x_0-\hat x||$ are sufficiently small. Such a priori
assumptions are typical for the methods of solving nonlinear
problems, in particular, for the Newton-type methods.

\noindent {\bf Remark 3.5} In practice, algorithm
(\ref{2.3})-(\ref{2.4}) proved to be efficient. In order to test
(\ref{2.3})-(\ref{2.4}) numerically, we considered a two-dimensional
inverse gravimetry problem (see \cite{ars1}, \cite{rs}) and the
Feigenbaum equation (see \cite{ars2}, equation (5.1)). The goal of
the experiments was to investigate the choice of the
regularization function $\ep(t)$ in (\ref{2.3})-(\ref{2.4}) and to
compare two continuous methods: (\ref{2.3})-(\ref{2.4}) and
(\ref{2.1}). Analyzing the results of practical computations we
arrived at the following conclusions:
\begin{enumerate}
  \item Continuously regularized Gauss-Newton's scheme
(\ref{2.1}) tends to do well initially, until the value of a
regularization function $\ep(t)$ is not very small. Continuous
procedure (\ref{2.3})-(\ref{2.4}) is more stable, and allows one
to get the approximate solution with better accuracy.
  \item The best numerical results for (\ref{2.3})-(\ref{2.4})
are obtained with $\ep(t)=\ep(0)(1+t)^{-1}$. An approximate range
of possible values for $\ep(0)$ appears to be from $0.001$ to
$0.1$; for larger values the accuracy is lower, and for smaller
values the process does not converge.
  \item An important task that should be studied next is the
stability of the solution towards the noise in the data, and the choice
of an optimal
regularization parameter (the stopping time in our case) such that
the method converges to the solution of (1.1) when  the noise level
tends to zero. Also one
has to analyze under what assumptions on $F$ and $x_0$ the
convergence of the continuous process (\ref{2.3})-(\ref{2.4})
implies the convergence of the corresponding discrete process,
obtained by the Euler or the Runge-Kutta finite difference
methods.
\end{enumerate}

{\bf Proof of Theorem~\ref{t1}.}
The main part of the proof (Steps 1-3) is to show that if
$(x(t),B(t))$ is a solution to (\ref{2.3})-(\ref{2.4}), then
$||x(t)-\hat x||<R\ep(t)$. Assume the converse. By (\ref{??})
one has the inequality
$||x(0)-\hat x||<R\ep(0).$ Therefore there exists a $\t\in
[0,+\infty)$, such that
\begin{equation}\label{^}
||x(t)-\hat x||<R\ep(t) \q \forall t\in [0,\t)\q \mbox{and} \q
||x(\t)-\hat x||=R\ep(\t).
\end{equation}

\noindent \underline{\bf Step 1.} In this step we derive some
estimates ( see (\ref{3.1})), (\ref{3.3}) and (\ref{3.4}) below)
for
$$||B(t)||, \quad
||I-B(t)\bigl[F^{\pr*}(\hat x)F'(\hat x)\\+\ep(t)I\bigr]||
\q\mbox { and }\q ||B(t)F^{\pr*}(\hat x)F'(\hat x)||.$$ We assume
below that $t\in [0,\t]$. Applying Lemma~\ref{l3}  with
$$
V(t):=B(t), \q A(t):=F^{\pr*}(x(t))F'(x(t))+\ep(t)I, \q G(t):=I,
$$
one gets
$$
||B(t)||\le
e^{-\il^t_0\ep(p)dp}\left[\il^t_0e^{\il^s_0\ep(p)dp}\,ds+||B(0)||\right]
\le
e^{-\il^t_0\ep(p)dp}\left[\f{1}{\ep(t)}\il^t_0\ep(s)e^{\il^s_0\ep(p)dp}\,ds+||B(0)||\right]
$$
\begin{equation}\label{3.1}
=\f{1}{\ep(t)}\left[1-e^{-\il^t_0\ep(p)dp}\right]+||B(0)||e^{-\il^t_0\ep(p)dp}\le\f{1}{\ep(t)}+||B(0)||.
\end{equation}
Introduce the notation
\begin{equation}\label{3.2}
\L(t):=I-B(t)\bigl[F^{\pr*}(\hat x)F'(\hat x)+\ep(t)I\bigr].
\end{equation}
Then
$$
\dot \L(t)=-\dot B(t)\bigl[F^{\pr*}(\hat x)F'(\hat
x)+\ep(t)I\bigr]-B(t)\dot \ep(t)
$$
$$
=\bigl[F^{\pr*}(x(t))F'(x(t))B(t)-I+\ep(t)B(t)\bigr]\bigl[F^{\pr*}(\hat
x)F'(\hat x)+\ep(t)I\bigr]-B(t)\dot \ep(t)
$$
$$
=\bigl[F^{\pr*}(x(t))F'(x(t))+\ep(t)I\bigr](I-\L(t))-\bigl[F^{\pr*}(\hat
x)F'(\hat x)+\ep(t)I\bigr]-B(t)\dot \ep(t)
$$
$$
=-\bigl[F^{\pr*}(x(t))F'(x(t))+\ep(t)I\bigr]\L(t)+
F^{\pr*}(x(t))F'(x(t))-F^{\pr*}(\hat x)F'(\hat x)-B(t)\dot \ep(t).
$$
Use Lemma~\ref{l3} once again, with
$$
V(t):=\L(t), \q A(t):=F^{\pr*}(x(t))F'(x(t))+\ep(t)I,
$$
$$
G(t):=F^{\pr*}(x(t))F'(x(t))-F^{\pr*}(\hat x)F'(\hat x)-B(t)\dot
\ep(t).
$$
From the estimate
$$
||F^{\pr*}(h)F'(h)-F^{\pr*}(y)F'(y)||=||F^{\pr*}(h)\bigl[F'(h)-F'(y)\bigr]+
\bigl[F^{\pr*}(h)-F^{\pr*}(y)\bigr]F'(y)||
$$
$$
\le 2N_1N_2||h-y|| \q \mbox{for any} \q h,y \in U(\hat x,R\ep(0)),
$$
and from (\ref{3.1}), (\ref{^}), (\ref{ep}) and condition 3) of
Theorem 3.1 one obtains:
$$
||\L(t)||\le e^{-\il^t_0\ep(p)dp} \left\{\il^t_0
\left[2N_1N_2||x(s)-\hat x||+\left (\frac{1}{\ep(s)}+\left(||B(0)||-\f{1}
{\ep(s)}\right)e^{-\il^s_0\ep(p)dp}\right)|\dot
\ep(s)|\right]\right.
$$
$$
\left.e^{\il^s_0\ep(p)dp}ds+||\L(0)||\right\}\le
e^{-\il^t_0\ep(p)dp}\left[(2N_1N_2R+b)\il^t_0\ep(s)
e^{\il^s_0\ep(p)dp}\,ds+\il_0^t\left(||B(0)||-\f{1}{\ep(s)}\right)
\right.
$$
$$
|\dot \ep(s)|ds+||\L(0)||\Biggr]=(2N_1N_2R+b)\left[1-
e^{-\il^t_0\ep(p)dp}\right]+\left(||B(0)||-\f{1}{\ep(0)}\right)(\ep(0)-\ep(t))+
$$
\begin{equation}\label{3.3}
+||\L(0)||e^{-\il^t_0\ep(p)dp} \le 2N_1N_2R+b+\ep(0)||B(0)||+||\L
(0)||:=k
\end{equation}
Now denote $D(t):=B(t)F^{\pr*}(\hat x)F'(\hat x)$. Then by
(\ref{3.3}) $||\L(t)||=||I-D(t)-B(t)\ep(t)||\le k,$ and therefore  by
(\ref{3.1})
\begin{equation}\label{3.4}
||D(t)||\le k+2+\ep(0)||B(0)||.
\end{equation}

\noindent \underline{\bf Step 2.} In this step we derive the
differential inequality (see (\ref{3.5}) below).

Since $H$ is a real Hilbert
space one has:
$$
\f{1}{2}\f{d}{dt}||x(t)-\hat x||^2=(\dot x,x(t)-\hat
x)=-(B(t)\bigl[F^{\pr*}(x(t))F(x(t))+\ep(t)(x(t)-x_0)\bigr],x(t)-\hat
x ).
$$
One gets
\begin{equation}\label{*}
F(x(t))=F'(\hat x)(x(t)-\hat x)+K(x(t),x(t)-\hat x, x(t)-\hat x),
\end{equation}
where
\begin{equation}\label{**}
||K(h,y,z)||\le \f{N_2}{2}||y||\,||z|| \q \mbox{for any} \q h\in
U(\hat x,R\ep(0)),\q y,z\in H.
\end{equation}
Thus, (\ref{*}), (\ref{**}) and (\ref{3.1}) yield
$$
\f{1}{2}\f{d}{dt}||x(t)-\hat x||^2\le -(B(t)F^{\pr*}(x(t))F'(\hat
x)(x(t)-\hat x),x(t)-\hat x)-\ep(t)(B(t)(x(t)-x_0),x(t)-\hat x)
$$
$$
+\f{N_1N_2(1+\ep(0)||B(0)||)}{2\ep(t)}||x(t)-\hat
x||^3=-(B(t)(F'(x(t))-F'(\hat x))^*F'(\hat x)(x(t)-\hat
x),x(t)-\hat x)
$$
$$
-(B(t)\bigl[F^{\pr*}(\hat x)F'(\hat x)+\ep(t)I\bigr](x(t)-\hat
x),x(t)-\hat x)-\ep(t)(B(t)(\hat x-x_0),x(t)-\hat
x)
$$
$$
+\f{N_1N_2(1+\ep(0)||B(0)||)}{2\ep(t)}||x(t)-\hat x||^3.
$$
From (\ref{3.2}) one gets
$$
-(B(t)\bigl[F^{\pr*}(\hat x)F'(\hat x)+\ep(t)I\bigr](x(t)-\hat
x),x(t)-\hat x)=-||x(t)-\hat x||^2
$$
\begin{equation}\label{***}
+(\L(t)(x(t)-\hat x),x(t)-\hat x).
\end{equation}
By estimate (\ref{3.1})
$$
-(B(t)(F'(x(t))-F'(\hat x))^*F'(\hat x)(x(t)-\hat x),x(t)-\hat x)
$$
\begin{equation}\label{****}
\le\f{N_1N_2(1+\ep(0)||B(0)||)}{\ep(t)}||x(t)-\hat x||^3.
\end{equation}
Condition 5) of Theorem~\ref{t1} implies:
\begin{equation}\label{*****}
-\ep(t)(B(t)(\hat x-x_0),x(t)-\hat x)=-\ep(t)(D(t)w,x(t)-\hat x).
\end{equation}
From (\ref{***}), (\ref{****}), (\ref{*****}), (\ref{3.3}),
(\ref{3.4})  one concludes
$$
\f{1}{2}\f{d}{dt}||x(t)-\hat
x||^2\le\f{3N_1N_2(1+\ep(0)||B(0)||)}{2\ep(t)}||x(t)-\hat x||^3
-(1-k)||x(t)-\hat x||^2
$$
$$
+\ep(t)(k+2+\ep(0)||B(0)||)\,||w||\,||x(t)-\hat x||.
$$
Let $v(t):=||x(t)-\hat x||$. Then we arrive at the inequality
\begin{equation}\label{3.5}
\dot
v\le-(1-k)v+\f{3N_1N_2(1+\ep(0)||B(0)||)}{2\ep(t)}v^2+\ep(t)(k+2+\ep(0)||B(0)||)\,||w||.
\end{equation}

\noindent \underline{\bf Step 3.} In this step we choose
$\mu(t)\to \infty$ as $t\to \infty,$ such that assumptions
(\ref{ric1}) of Lemma~\ref{l2} are satisfied for inequality
(\ref{3.5}), so that the conclusion $||x(t)-\hat x||<\frac 1
{\mu(t)}$ holds.

Choose $\mu(t)\in
C^1[0,+\infty)$ to satisfy conditions (\ref{ric1}) of
Lemma~\ref{l2} with
\begin{equation}\label{3.6-3.7}
\g(t):=1-k,\q \s(t):=\f{3N_1N_2(1+\ep(0)||B(0)||)}{2\ep(t)},
\end{equation}
\begin{equation}\label{3.8}
\b(t):=\ep(t)(k+2+\ep(0)||B(0)||)\,||w||.
\end{equation}
Take $\mu(t)=\f{\l}{\ep(t)},$
where $\lambda>0$ is a constant. Then (\ref{ric1}) can be rewritten
as follows:
\begin{equation}\label{3.9}
\f{3N_1N_2(1+\ep(0)||B(0)||)}{2\ep(t)}\le
\f{\l}{2\ep(t)}\left(1-k-\f{|\dot \ep(t)|}{\ep(t)}\right),
\end{equation}
\begin{equation}\label{3.10}
\ep(t)(k+2+\ep(0)||B(0)||)\,||w||\le
\f{\ep(t)}{2\l}\left(1-k-\f{|\dot \ep(t)|}{\ep(t)}\right),
\end{equation}
\begin{equation}\label{3.11}
\f{\l}{\ep(0)}||x(0)-\hat x||<1.
\end{equation}
Inequality (\ref{3.9}) is equivalent to the following one
\begin{equation}\label{3.12}
\l\ge\f{3N_1N_2(1+\ep(0)||B(0)||)}{1-k-\f{|\dot \ep(t)|}{\ep(t)}}.
\end{equation}
If one takes
\begin{equation}\label{3.13}
\l:=\f{3N_1N_2(1+\ep(0)||B(0)||)}{1-k-b\ep(0)},
\end{equation}
then (\ref{3.12}) follows from (\ref{3.13}) and from conditions
1) and 4) of Theorem~\ref{t1}. Assumption (\ref{??}) of
Theorem~\ref{t1} implies that
\begin{equation}\label{3.14}
\f{3N_1N_2(1+\ep(0)||B(0)||)}{1-k-b\ep(0)}<\f{1-k-b\ep(0)}{2(k+2+\ep(0)||B(0)||)||w||}.
\end{equation}
For $\l$ defined by (\ref{3.13}) inequality (\ref{3.14}) can be
written as
\begin{equation}\label{3.15}
\l<\f{1-k-b\ep(0)}{2(k+2+\ep(0)||B(0)||)||w||}.
\end{equation}
From (\ref{3.15}) one obtains (\ref{3.10}).

Finally, from
(\ref{??}) one concludes that
\begin{equation}\label{3.16}
\f{3N_1N_2(1+\ep(0)||B(0)||)}{1-k-b\ep(0)}<\f{\ep(0)}{||x(0)-\hat
x||},
\end{equation}
which is equivalent to (\ref{3.11}) for $\l$ defined by
(\ref{3.13}).

Hence, by Lemma~\ref{l2}, one obtains:
$$||x(t)-\hat x||<\f{\ep(t)}{\l}.$$
By (\ref{3.13}) and (\ref{??})
$$
\f{||x(t)-\hat
x||}{\ep(t)}<\f{1-k-b\ep(0)}{3N_1N_2(1+\ep(0)||B(0)||)}\le R \q
\mbox{for all} \q t\in [0,\t],
$$
which contradicts (\ref{^}). Thus $||x(t)-\hat x||<R\ep(t)$
whenever $x(t)$ is defined.

\noindent \underline{\bf Step 4.} Now let us show that there
exists the unique solution to (\ref{2.3})-(\ref{2.4}) on
$[0,+\infty)$. The initial value problem (\ref{2.3})-(\ref{2.4})
is equivalent to the integral equations:
\begin{equation}\label{3.x}
x(t)=x_0+\il_0^tB(s)\P(s,x(s))ds,\q x_0\in H,
\end{equation}
\begin{equation}\label{3.b}
B(t)=B(0)+\il_0^t[\p(s,x(s))B(s)+I]ds,\q B(0)\in L(H),
\end{equation}
where
$$
\P(s,x):=-[F^{\pr*}(x)F(x)+\ep(s)(x-x_0)],
$$
$$
\p(s,x):=-[F^{\pr*}(x)F'(x)+\ep(s)I].
$$
We have already proved that if for some positive number $T$ the
solution $(x(t),B(t))$ to problem (\ref{2.3})-(\ref{2.4}) exists
on $[0,T]$, then $x(t)$ and $B(t)$ are bounded in the norm on
$[0,T]$.  Therefore, since the integrands in (\ref{3.x}) and
(\ref{3.b}) are Lipschitz-continuous, the standard argument yields
existence and uniqueness of the global solution to system
(\ref{3.x})-(\ref{3.b}). This concludes Step 4.

For convenience of the reader we give a complete proof of Step 4.
Let us fix an arbitrary large positive number $T$ and use the
successive approximation method to solve equations
(\ref{3.x})-(\ref{3.b}) on $[0,T]$:
\begin{equation}\label{3.xn}
x_{n+1}(t)=x_0+\il_0^tB_n(s)\P(s,x_n(s))ds,\q x_0\in H,
\end{equation}
\begin{equation}\label{3.bn}
B_{n+1}(t)=B(0)+\il_0^t[\p(s,x_n(s))B_n(s)+I]ds,\q B(0)\in L(H).
\end{equation}
Since $F'(x)$ and $F''(x)$ are assumed to be bounded in $U(\hat
x,R\ep(0))$, and $\ep(t)$ is positive on $[0,T]$, for every $t\in
[0,T]$ the function $\P(s,x)$ has bounded Fr\'echet derivative
with respect to $x$ in $U(\hat x,R\ep(0))$. So one has:
$$
||B_2\P(s,x_2)-B_1\P(s,x_1)||=||(B_2-B_1)\P(s,x_2)+B_1(\P(s,x_2)-\P(s,x_1))||
$$
$$
\le A(T)(||B_2-B_1||+||x_2-x_1||)
$$
for all $t\in [0,T]$ and $x_1,x_2\in U(\hat x,R\ep(0))$. Similarly,
$$
||\p(s,x_2)B_2-\p(s,x_1)B_1||=||(\p(s,x_2)-\p(s,x_1))B_2+\p(s,x_1)(B_2-B_1)||
$$
$$
\le K(T)(||B_2-B_1||+||x_2-x_1||).
$$
Define the norm in a Banach space $H\times L(H)$ as follows:
$$
||(x,B)||:=||x||+||B||.
$$
Thus, if $w_i(t):=(x_i(t),B_i(t))$, $i=1,2,...$, one gets the
estimate
$$
||w_{n+1}(t)-w_n(t)||\le \il _0^t
||B_n(s)\P(s,x_n(s))-B_{n-1}(s)\P(s,x_{n-1}(s))||\,ds
$$
$$
+\il_0^t||\p(s,x_n(s))B_n(s)-\p(s,x_{n-1}(s))B_{n-1}(s)||\,ds
$$
$$
\le \il^t_0 (A(T)+K(T))||w_n(s)-w_{n-1}(s)||\,ds\,\le ... \le\,
\f{((A(T)+K(T))T)^n}{n!} \, const
$$
valid on the maximal subinterval $[0,T_{max}]:=\{t,\,t\in
[0,T]\,\,\mbox{and}\,\,x(t)\in U(\hat x, R\ep(0))\}$. Therefore
iterative process (\ref{3.xn})-(\ref{3.bn}) converges uniformly
and determines the unique solution $(x(t),B(t))$ of problem
(\ref{3.x})-(\ref{3.b}) on $[0,T_{max}]$. If $T_{max}<T$, it
follows from the maximality of the subinterval $[0,T_{max}]$ that
$x(T_{max})$ is a boundary point of $U(\hat x, R\ep(0))$. But,
since $\ep(t)$ is monotone, this contradicts the fact that the
inequality  $||x(t)-\hat x||<R\ep(t)$ holds whenever $x(t)$ is
defined. Hence the unique solution $(x(t),B(t))$ to
(\ref{2.3})-(\ref{2.4}) exists on $[0,T]$ for any $T>0$.
Consequently, it exists on $[0,+\infty)$ and (\ref{lim}) is
satisfied $\forall t\in [0,+\infty)$. Theorem~\ref{t1} is proved.
$\qed$

\end{document}